\documentclass[final,3p,times,twocolumn]{elsarticle}
 \biboptions{comma,sort&compress}
\usepackage{here}
 \usepackage{graphicx}
  \usepackage{epsfig}
\usepackage{graphics,psfrag}

\usepackage{amsmath}
\usepackage{amssymb}

\def\nin{\noindent}
\def\beq{\begin{equation}}
\def\eeq{\end{equation}}
\def\bea{\begin{eqnarray}}
\def\eea{\end{eqnarray}}

\newcommand{\ep}{\epsilon}
\newcommand{\nn}{\nonumber}
\newcommand{\vp}{\mathbf{p}}

\journal{Nuc. Phys. (Proc. Suppl.)}

\begin{document}

\begin{frontmatter}



\title{Hadron resonances generated from the dynamics of the lightest scalar ones}

 \author[label1]{J.~A.~Oller\corref{cor1}}
  \address[label1]{Departamento de F\'{\i}sica, Universidad de Murcia, E-30071 Murcia, Spain}
\cortext[cor1]{Speaker}
\ead{oller@um.es}

\author[label1]{J.~M.~Alarc\'on}
\author[label1]{ M.~Albaladejo}
 \author[label2]{L.~Alvarez-Ruso}
 \author[label1]{L.~Roca}
  \address[label2]{Centro de F\'{\i}sica Computacional, Departamento de F\'{\i}sica, Universidade de Coimbra, Portugal
}


\begin{abstract}
\noindent
We have studied the interactions of the scalar resonances $f_0(980)$ and $a_0(980)$ with the vector resonance $\phi(1020)$ and 
with the lightest pseudoscalars $\pi$, $K$, $\eta$ and $\eta'$.  We first obtain the interaction kernels without including any new free parameter. Afterwards, the interaction kernels are unitarized and the final S-wave amplitudes result. 
We find that these interactions are very rich and generate a large amount of pseudoscalar resonances including the 
 $K(1460)$, $\pi(1300)$, $\pi(1800)$, $\eta(1475)$ and $X(1835)$ resonances. The $f_0(980)\phi(1020)$ self-interactions 
 give rise to the $\phi(2170)$ resonance.  For realistic choices of the parameters we also obtain  an isovector companion in the same mass region from the 
$a_0(980) \phi(1020)$ interactions.

\end{abstract}


\end{frontmatter}


\section{Introduction}
\nin

Due to the spontaneous chiral symmetry breaking of QCD the interactions between the lightest pseudoscalars are constrained   \cite{wein,gas1}.
 These interactions in S-wave are strong enough to  generate dynamically the lightest scalar resonances, 
namely, the $f_0(980)$, $a_0(980)$, 
$\kappa$ and $\sigma$, as shown in refs.~\cite{npa,doba,iamprl,nd}. 
Still one can make use of the tightly constrained interactions among the lightest pseudoscalars in order to work out
 approximately the scattering between the lightest pseudoscalar mesons and scalar resonances. 
We report here about ref.~\cite{alba2} where the narrower resonances $f_0(980)$ and $a_0(980)$ are taken and 
 the scattering of the latter ones with the pseudoscalars $\pi$, $K$, $\eta$ and $\eta'$ is studied. Many pseudoscalar resonances 
then arise as  dynamically generated. 
The problem of the excited pseudoscalars above 1~GeV is interesting by itself. These resonances are not well-known typically \cite{pdg}. In isospin $I=1/2$ one has 
the $K(1460)$ and $K(1630)$ resonances.
 The broad $I=1$ resonances $\pi(1300)$, $\pi(1800)$ are  somewhat better known \cite{pdg}. 
Special mention deserves the $I=0$ channel where the $\eta(1295)$, 
$\eta(1405)$, $\eta(1475)$ have been object of an intense theoretical and 
experimental study \cite{masoni_rev}.   Refs.~\cite{pdg,masoni_rev} favor the     description of  the  
$\eta(1295)$ and $\eta(1405)$ as ideally mixed states  of the same nonet of pseudoscalar resonances, with the other members being the $\pi(1300)$ and the $K(1460)$.
The $\eta(1405)$ would then be an extra state  whose clear signal in gluon-rich processes would favor its interpretation as a 
glueball in QCD \cite{chano,close}. 
However, this conclusion clashes with lattice QCD that 
predicts the lowest mass for the pseudoscalar glueball  around 2.4~GeV \cite{bali}.
 Nonetheless, the $\eta(1405)$ would fit as a $0^{-+}$ glueball, if the latter is 
 a closed gluonic fluxtube  \cite{faddev}. Its mass and properties also fit as  a gluino-gluino bound state \cite{17,close}. 
 The previous whole picture for classifying the lightest pseudoscalar resonances has been challenged in ref.~\cite{klempt}. 
 On the other hand, the  resonance $X(1835)$  was recently observed by the BES Collaboration and the assignment of pseudoscalar quantum numbers $0^{-+}$  was favored \cite{ablikim}. 

The  resonance $\phi (2170)$ (also denoted by $Y(2175)$) cannot be easily accommodated within the quark model \cite{Zhu:2007wz}. It was first observed by the BABAR Collaboration~\cite{babar1}. 
 Ref.~\cite{Shen:2009mr} obtained $M_Y=2117^{+0.59}_{-0.49}$~MeV and $\Gamma_Y=164^{+69}_{-80}$~MeV from a combined fit to both BABAR and Belle data. 
In Ref.~\cite{fif0} our research group  achieved a good description of the $e^+e^-\to \phi(1020) f_0(980)$  data in the threshold region ($\sim 2$~GeV) using chiral
 Lagrangians coupled to vector mesons, supporting the conclusion that the $\phi(2170)$ has a large $\phi(1020) f_0(980)$ mesonic component \cite{torres}. This study was extended 
in ref.~\cite{alva} to consider the $I=1$ $a_0(980)\phi(1020)$ scattering so as to disentangle whether an isovector companion of the $\phi(2170)$
 could appear \cite{mexico}.

\section{Formalism. Setting the model}
\label{sec:form}
\nin
In order to settle the formalism we consider first the interactions between the scalar resonances and the lightest pseudoscalars 
following ref.~\cite{alba2}. The interactions of the former resonances with the $\phi(1020)$ will emerge as a special case.

\begin{figure}[ht]
\psfrag{p1}{{\small $p_1$}}
\psfrag{p2}{{\small $p_2$}}
\psfrag{S1}{{\small $ S_1 $}}
\psfrag{S2}{{\small $ S_2 $}}
\psfrag{l1}{{\small $\ell$}}
\psfrag{l2}{{\small $p_1-\ell$}}
\psfrag{l3}{{\small $p_2-\ell$}}
\psfrag{k1}{{\small $k_1$}}
\psfrag{k2}{{\small $k_2$}}
\psfrag{P1}{{\small $\begin{array}{c} \\ P_1 \end{array}$}}
\psfrag{P2}{{\small $\begin{array}{c} \\ P_2 \end{array}$}}
\centerline{\epsfig{file=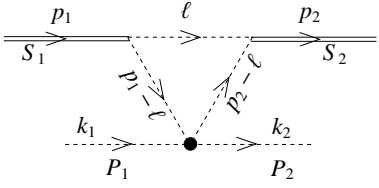,width=.3\textwidth,angle=0}}
\vspace{0.2cm}
\caption[pilf]{\protect \small
Triangular loop for calculating the interaction kernel for $S_1(p_1)P_1(k_1)\to S_2(p_2)P_2(k_2)$. 
\label{fig:tri}}
\end{figure}

Our approach is based on the triangular diagram shown in fig.~\ref{fig:tri} where an 
incident scalar resonance $S_1$ decays into a virtual $K\bar{K}$ pair.
 The filled dot in the vertex on the 
bottom of the diagram corresponds to the interaction of the incident (anti)kaon in the loop with 
the pseudoscalar $P_1$ giving rise to the the pseudoscalar $P_2$ and the same (anti)kaon. This scattering amplitude is denoted by  $T((P-\ell)^2)$, 
 with $P=p_1+k_1$ the total four-momentum and $\ell$ the running one in the loop. 
The out-going scalar resonance is denoted by $S_2$.  
The basic point is that this diagram is enhanced because the masses
 of  both  the $f_0(980)$ and the $a_0(980)$ resonances 
 are very close to the $K\bar{K}$ threshold. 
In this way, for  scattering near the threshold of the reaction, one of the 
kaon lines in the bottom of the diagram is almost on-shell. Indeed, at threshold and in the limit of the mass of the scalar equal to twice the kaon mass this diagram becomes infinite \cite{fif0}. 
 One takes advantage of the fact that both the 
$f_0(980)$ and $a_0(980)$ resonances are dynamically generated by the 
meson-meson self-interactions \cite{npa,nd,isgur}.  In this way, we can calculate the couplings of the scalar resonances considered 
to two pseudoscalars, including their relative phase.
 The coupling of the $f_0(980)$ and $a_0(980)$ resonances to a $K\bar{K}$ pair in $I=0$ and 1, respectively, is denoted by $g_{f_0}$ and $g_{a_0}$.
  The diagram in fig.~\ref{fig:tri} reads
 {\small \begin{align}
 L_K&= i \int\frac{d^4\ell}{(2\pi)^4}\frac{T((P-\ell)^2)}{(\ell^2-m_K^2+i\ep)((p_1-\ell)^2-m_K^2+i\ep)}
\nn\\
&\times\frac{1}{ ((p_2-\ell)^2-m_K^2+i\ep)}~.
 \label{tl.ref}
 \end{align}}
 Here, we employ 
the meson-meson scattering amplitudes obtained in ref.~\cite{nd} but now enlarged,
so that states with the pseudoscalar  $\eta'$ are included. These amplitudes contain the poles of the scalar resonances $\sigma$, $\kappa$, 
$f_0(980)$, $a_0(980)$ and other poles in the region around 1.4~GeV \cite{nd}.
 In order to proceed further we have to know the dependence of $T((P-\ell)^2)$ on  the integration variable $\ell$.
 This can be done by writing the dispersion relation satisfied by $T(q^2)$, 
{\small \begin{align}
T(q^2)&=T(s_A)+\sum_i\frac{q^2-s_A}{q^2-s_i}\frac{\hbox{Res}_i}{s_i-s_A}\nn\\
&+\frac{q^2-s_A}{\pi}\int_{s_{th}}^\infty ds'\frac{\hbox{Im}T(s')}{(s'-q^2)(s'-s_A)}~.
\label{dis.rel}
\end{align}} 
One subtraction at $s_A$ has been taken because $T(q^2)$ is bounded by a constant 
for $q^2\to\infty$, with $T(s_A)$ the subtraction constant. Typically there are also present 
deep poles in the $s$-complex plane  at $s_i$ whose residue are Res$_i$.  
 Inserting eq.~\eqref{dis.rel} 
into eq.~\eqref{tl.ref} one can write for $L_K$
{\small \begin{align}
L_K&=\left(T(s_A)+\sum_i\frac{\hbox{Res}_i}{s_i-s_A}\right)C_3
+\sum_i C_4(s_i) \hbox{Res}_i \nn\\
&-\frac{1}{\pi}\int_{s_{th}}^\infty ds' \hbox{Im}T(s') 
\left[\frac{C_3}{s'-s_A}+C_4(s') \right]~.
\label{tl.dis}
\end{align}} 
Here we have introduced the three- and four-point Green functions $C_3$ and $C_4(M_4^2)$ defined in standard 
way
{\small 
\begin{align}
C_3&=i\int\frac{d^4\ell}{(2\pi)^4}
\frac{1}{(\ell^2-m_K^2+i\ep)((p_1-\ell)^2-m_K^2+i\ep)}\nn\\
&\times \frac{1}{((p_2-\ell)^2-m_K^2+i\ep)}~,\nn\\
 C_4(M_4^2)&=i\int\frac{d^4 \ell}{(2\pi)^4} \frac{1}{(\ell^2-m_K^2+i\ep)((p_1-\ell)^2-m_K^2+i\ep)}\nn\\
&\times \frac{1}{((p_2-\ell)^2-m_K^2+i\ep)((P-\ell)^2-M_4^2+i\epsilon)}~.
 \label{c3.c4.def}
\end{align}}
\normalsize
 One has still to perform the 
angular projection for $C_3$ and $C_4(M_4^2)$, see appendix B of ref.~\cite{alba2}. Once this is done eq.~\eqref{tl.dis} can 
still be used but with $C_3$ and $C_4(M_4^2)$ projected in S-wave.
 For $S_1(p_1) P_1(k_1)\to S_2(p_2) P_2(k_2)$ we have the usual Mandelstam variables 
$s=(p_1+k_1)^2$, $t=(p_1-p_2)^2$ and $u=(p_1-k_2)^2=
M_{S_1}^2+M_{S_2}^2+M_{P_1}^2+M_{P_2}^2-s-t$, with the masses of the particles indicated by $M$ 
with the subscript distinguishing between them. The dependence on the relative angle 
$\theta$  enters in $t$ as 
$t=(p_1^0-p_2^0)^2-(\vp-\vp')^2=(p_1^0-p_2^0)^2-\vp^2-\vp'^2+2|\vp||\vp'|\cos\theta$ with $\vp$ and 
$\vp'$ the CM three-momentum of the initial and final particles, respectively.

 Eq.~\eqref{tl.dis} is our basic equation for evaluating the interaction kernels. One has 
 only to specify the pseudoscalars actually involved in the amplitude $T((P-\ell)^2)$,  according to the 
 specific reaction under consideration.  
For each set of quantum numbers, specified by the isospin $I$ and G-parity $G$ (if the latter is not defined this label should be omitted), we join 
in a symmetric matrix ${\cal T}_{IG}$ the different 
 $T_L(i\to j)$ calculated. Then, in order to resum the unitarity loops  and obtain  the final  S-wave scalar-pseudoscalar T-matrix, $T_{IG}$, 
 we make use of the master equation
{\small  \begin{align}
T_{IG}&=\left[I+{\cal T}_{IG}\cdot g_{IG}(s)\right]^{-1}\cdot {\cal T}_{IG}~.
\label{t.uni}
\end{align}}
For a general derivation of this equation, based on the N/D method see refs.~\cite{nd,plb,npa}.
 In eq.~\eqref{t.uni} $g_{IG}(s)$ is a diagonal matrix whose elements are 
the scalar unitarity loop function with a scalar-pseudoscalar intermediate state. 
For the calculation of $g_{IG}(s)_i$, corresponding to  the $i_{th}$ 
state with the quantum numbers $IG$ and made up by  the scalar resonance 
$S_i$ and the pseudoscalar $P_i$, we make use of a once subtracted dispersion relation. Explicit expressions 
are given in refs.~\cite{nd,fif0,alva}.  The subtraction constant $a_1$ 
is restricted to have natural values so that the unitarity scale \cite{fif0} $4\pi f_\pi/\sqrt{|a_1|}$
 becomes not too small (e.g. below the $\rho$-mass), $|a_1|\lesssim 3$. In addition,
 we require the sign of $a_1$ to be negative so that resonances could be generated when the interaction kernel is positive (attractive).  


For the interactions of the considered scalar resonances with the $\phi(1020)$ we take into account fig.~\ref{fig:tri}. In this case, the horizontal dashed lines at the bottom correspond to the $\phi(1020)$ and its interaction to $K\bar{K}$ is obtained by employing the lowest order chiral Lagrangians coupled to vector mesons by minimal coupling. It is then calculated at tree level and it is characterized by a coupling $g^2$. The interaction kernel is then proportional to  $g^2\, C_3$. Together with fig.~\ref{fig:tri}, local terms are also introduced coupling directly 
the $\phi(1020)$ with two kaons. The unitarization of the resulting kernels is performed employing eq.~\eqref{t.uni} with the proper masses for the resonances involved.    For further details we refer to refs.~\cite{fif0,alva}. 

\section{Results}
\nin

\begin{figure}[ht]
\psfrag{X(1835)}{{\tiny $X(1835)$}}
\psfrag{e(1475)}{{\tiny $\eta(1475)$}}
\psfrag{ss}{{\tiny $\sqrt{s}$ [MeV]}}
\psfrag{TS(f0eptof0ep)}{{\tiny $|T(f_0 \eta'\to f_0 \eta')|^2$}}
\psfrag{TS(f0etof0e)}{{\tiny $|T(f_0 \eta\to f_0 \eta)|^2$}}
\centerline{\epsfig{file=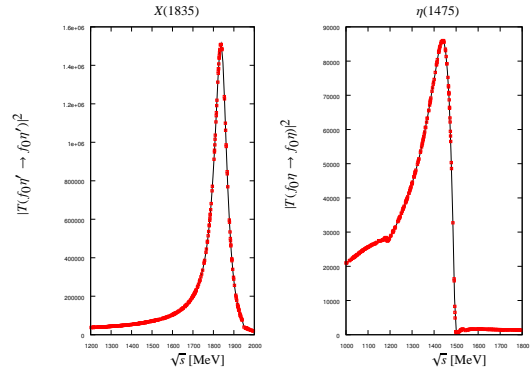,width=.3\textwidth,,keepaspectratio,angle=-90}}
\vspace{0.2cm}
\caption[pilf]{\protect \small
Modulus squared of the $f_0 \eta'\to f_0 \eta'$ (left)  and 
$f_0 \eta\to f_0 \eta$ (right) S-wave amplitudes.
\label{fig:i0}}
\end{figure}

The channels coupled for each quantum numbers are the $a_0\pi$, $f_0\eta$ and $f_0\eta'$ 
states for $I=0$, $G=+1$,   $f_0 K$ and $a_0 K$ in $I=1/2$ and $f_0\pi$, $a_0\eta$ and $a_0\eta'$ in $I=1$, $G=-1$. For the  exotic 
quantum numbers  $I=1$, $G=+1$ and $I=3/2$ one has the channels $a_0\pi$ and $a_0 K$, respectively. 
We obtain resonant signals for the $I=0$  resonances $\eta(1475)$ and $X(1835)$, for the $I=1/2$ resonance  $K(1460)$ and for the 
 $I=1$ $G=-1$  $\pi(1300)$ and $\pi(1800)$. For the exotic channel with $I=3/2$ we can also obtain a resonance for lower values of the subtraction constant, in agreement with the prediction of ref.~\cite{longa}.  We show in fig.~\ref{fig:i0} the resonant peaks corresponding to the 
$X(1835)$, left panel, and $\eta(1475)$, right panel. The $f_0\eta'$ channel, due to its high threshold, almost decouples from the $f_0\eta$ and $a_0\pi$ channels and produces a resonance  whose mass and width is in agreement with that of the $X(1835)$  \cite{pdg} for $a_1\simeq -1.2$. 
We also show in the right panel of the figure the modulus squared of the $f_0\eta$ scattering amplitude showing a clear resonance at 1475~MeV for $|a_1|\geq 0.8$. This resonance is associated with the $\eta(1475)$ because it does not couple to the $a_0(980)\pi$, the main decay channel of the nearby $\eta(1405)$. The $a_0\eta'$ channel almost decouples in the $\pi$-like channel with $I=1$ $G=-1$ (similarly as for the $f_0\eta'$ case) and gives rise to a resonant signal at the $ \pi(1800)$ resonance. We also obtained a clear signal for the $\pi(1300)$ resonance. For some resonances, like the $\eta(1475)$ or $K(1460)$, the influence of the channels involving the $K^*(892)$ resonance is expected to be significant because this is an important decay channel. We expect in the future to study 
simultaneously the S-wave scalar-pseudoscalar interactions together with the P-wave  vector-pseudoscalar ones.

\begin{figure}[ht]
\psfrag{nb}{{\tiny $\sigma$ [nb]}}
\psfrag{GeV}{{\tiny $\begin{array}{c}
\\
\sqrt{s} \,\hbox{[GeV]}
\end{array}$}}
\psfrag{sig para fi(1020)f0(980) g2 negativa ahora}{$\sigma(e^+e^-\to \phi(1020)f_0(980))$}
\centerline{\epsfig{file=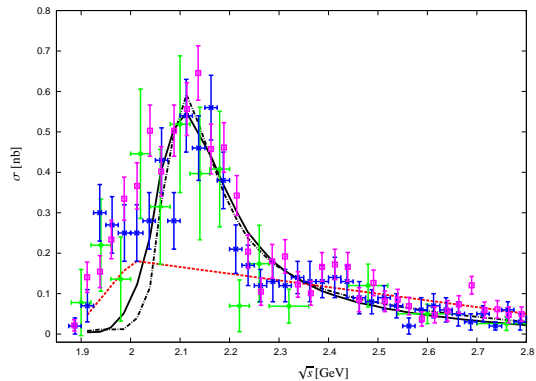,width=.3\textwidth,angle=-90}}
\vspace{0.2cm}
\caption[pilf]{\protect {\small
Cross-section for $e^+e^-\to \phi(1020)f_0(980)$. The experimental data are from Ref.~\cite{babar1} (diamonds and crosses) and Ref.~\cite{belle} (empty boxes). The dashed line shows  the non-resonant cross section from ref.~\cite{belle}.}
\label{fig:f0fi}}
\end{figure} 

\begin{figure}[ht]
\begin{center}
\psfrag{sigma (nb)}[Bb]{{\tiny $\sigma$~[nb]}}
\psfrag{sqs}[tc]{{\tiny $\sqrt{s}$~[GeV]}}
\centerline{\epsfig{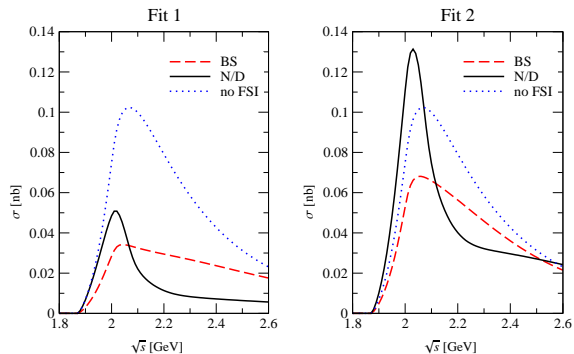}}
\vspace{0.2cm}
\caption[pilf]{\protect \small (Color online). $e^+e^-\to \phi(1020)\pi^0\eta$ cross section in the $a_0(980)$ region 
($M_{\pi\eta}\in[0.85,1.10]$~GeV).  
\label{fig:a0fi}}
\end{center}
\end{figure} 
For the case of  the $f_0(980)\phi(1020)$ we show in fig.~\ref{fig:f0fi} the fit performed in ref.~\cite{fif0} to the 
cross section  $\sigma(e^+ e^-\to\phi(1020)f_0(980))$. This fit employs the resulting S-wave $f_0(980)\phi(1020)$  amplitude to correct 
by final state interactions the non-resonant cross section of ref.~\cite{belle}. The agreement is very good. The two curves shown in the figure 
correspond to different values of the $f_0(980)$ mass, 0.98~GeV (solid line or Fit 1) and 0.99~GeV (dashed line or Fit 2). Finally, regarding the $a_0(980)\phi(1020)$ 
scattering we show  in fig.~\ref{fig:a0fi} the similar cross section $\sigma(e^+ e^-\to a_0(980)\phi(1020))$. The 
dotted lines correspond to the non-resonant cross section from ref.~\cite{mexico}. The latter has been dressed by final state interactions due to the $a_0(980)\phi(1020)$ S-wave  scattering amplitude giving rise to the solid and dashed lines. The solid lines  employ the 
$a_0(980)$ pole from ref.~\cite{nd}, while the dashed lines correspond to the $a_0(980)$ pole of ref.~\cite{npa}. The differences between the panels arise due to the fit employed for $\sigma(e^+e^-\to f_0(980)\phi(1020)$), as just discussed. One clearly sees a resonant behavior when the $a_0(980)$ pole is taken from 
the more complete study of the meson-meson amplitudes of ref.~\cite{nd} as compared with ref.~\cite{npa}, for which no resonance stems.

\section{Conclusions}
\nin
We have shown that the interactions of the scalar resonances with the lightest pseudoscalars and vector resonances is tightly constrained from chiral symmetry. 
Chiral Lagrangians are first used to generate dynamically the lightest nonet of scalar resonances and then a tower of new pseudoscalar and vector resonances 
can be generated dynamically by working out the interactions of the former with the pseudoscalar mesons and vector resonances employing again the same Lagrangians 
and the obtained meson-meson amplitudes containing the lightest scalars.  We have obtained resonant signals 
contributing to the physical pseudoscalar resonances $\pi(1300)$, $\pi(1800)$, $K(1460)$, $\eta(1475)$ and $X(1835)$. In addition, our solutions could also contain 
an exotic pseudoscalar $I=3/2$ $a_0 K$ resonance. We have obtained a good reproduction of the 
observed properties of the $\Phi(2170)$ as well and stressed the interest in looking for an isovector partner of the latter which is rather likely to exist as we have shown.









\end{document}